\documentclass{iau}
\usepackage{graphicx}

\title[Str\"omgren photometry and  spectroscopy  
] 
{Str\"omgren photometry and medium-resolution spectroscopy of some $\delta$~Scuti and
$\gamma$~Doradus in the \textit{Kepler} field }

\author[L. Fox-Machado]   
{L. Fox-Machado}

\affiliation{Instituto de Astronom\'{\i}a, Universidad Nacional Aut\'onoma de M\'exico\\ email: {\tt lfox@astrosen.unam.mx}}

\pubyear{2013}
\volume{301}  
\pagerange{1--2}
\jname{Precision Asteroseismology}
\editors{J.A. Guzik, W.J. Chaplin, G. Handler \& A. Pigulski, eds.}
\begin{document}

\maketitle

\begin{abstract}
We have obtained CCD photometry and medium-resolution spectroscopy of a number of $\delta$~Scuti
and $\gamma$~Doradus stars in the Kepler field-of-view as part of the ground-based observational 
efforts to support the \textit{Kepler} space mission. In this work we present the preliminary results
of these observations.
\end{abstract}

\firstsection 
\section{Introduction}

The \textit{Kepler} space mission (\cite[Borucki et al. 2010]{borucki}) was successfully  launched in March
2009 and since then it has been monitoring  a huge number of stars 
 in a region of 105  square degrees located between the constellations 
of Cygnus and Lyra. Although the main scientific goal of the mission is to discover 
Earth-sized planets,  the high precision 
photometry provided by the \textit{Kepler} satellite gives a unique opportunity to study
the pulsational variability of thousands of stars across the HR diagram in details
by means of asteroseismic methods (\cite[Aerts et al. 2010]{aerts}). As is well known, asteroseismic studies
require accurate and precise atmospheric parameters of the stars to produce reliable results. 
Since the precision of the physical parameters like effective temperature, gravity and metallicity
available in the \textit{Kepler Input Catalog} (KIC, Latham et al. 2005) is
 generally too low for asteroseismic modelling,
to best exploit the \textit{Kepler} data additional multi-colour and spectroscopic information is needed.
 In the framework of the \textit{Kepler} Asteroseismic Science Consortium
(KASC, \textit{http://astro.phys.au.dk/KASC/}) several ground-based observational efforts have been undertaken to derive
physical parameters of the \textit{Kepler} stars with high precision (e.g.~\cite[Uytterhoeven et al.~2011]{uytter}, \cite[Molenda-\.Zakowicz et al.~2011]{molenda}).
This paper describes our observational efforts at the
Observatorio Astron\'omico Nacional at San Pedro M\'artir (OAN-SPM) in Baja California, Mexico 
to derive the physical parameters of several $\delta$ Scuti and $\gamma$ Doradus stars in the \textit{Kepler}
field.

 \section{Observations, data reduction and conclusion}

The CCD  observations  of 74  $\delta$ Scuti and  $\gamma$ Doradus
stars in the \textit{Kepler} field  have been made with the 0.84-m
f/15 Ritchey-Chr\'etien telescope at OAN-SPM,  
during  six consecutive nights, from 2012 June 21 to June 26. The 
telescope hosted the filter-wheel `Mexman' with the ESOPO (E2V) CCD camera,
which has a 2048 $\times$ 4608 pixel array, with a pixel size of 15 $\times$ 15 $\mu$m$^{2}$.
The typical field-of-view with this configuration is 8$^{\prime}$ $\times$ 8$^{\prime}$.  The observations were taken
with Str\"omgren $uvby$ and H$\beta$ filters to take advantage of 
the  Str\"omgren-Crawford photometric system in deriving physical parameters of the stars.  
A set of standard stars from well observed open clusters (e.g. \cite[Pe\~na et al. 2011]{pena}) was also observed 
each night to transform instrumental observations
onto the standard system and to correct for atmospheric extinction. The usual calibration procedures for CCD photometry have been
carried out during  our observing run. Sky flat fields, bias and dark exposures were taken every night.
 The data reduction of this CCD photometry has been
carried out with the usual techniques and IRAF packages.
The instrumental magnitudes and colours, once corrected for atmospheric extinction were transformed to
the standard system.
In this way, we have obtained the standard Str\"omgren indices not only of the 74 $\delta$ Scuti and $\gamma$ Doradus
target stars, but also of all the stars brighter than V $\sim$ 15 mag located within each observed field.

\begin{figure}[t]
\begin{center}
 \includegraphics[width=3.4in]{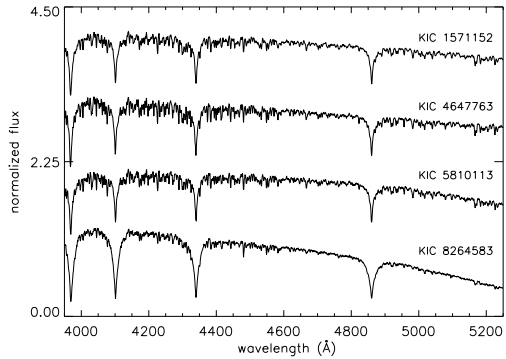} 
 \caption{Reduced spectra of some \textit{Kepler} targets.}
   \label{fig1}
\end{center}
\end{figure}

The spectroscopic observations  were conducted at the 2.12-m
telescope of the OAN-SPM observatory during several short runs
between 2010 and 2012. We used  the same equipment as explained in \cite[Fox Machado et al.~(2010)]{fox}.
In particular, we used 
the Boller \& Chivens spectrograph installed in the Cassegrain focus of the telescope. 
The 1200 lines/mm grating with a blaze angle of 13$^{\circ}$ was used.
The grating angle was set to 19$^{\circ}$ to cover a wavelength range from 3950~{\AA} to 5250~{\AA}.
A 2048$\times$2048 E2V CCD camera
was used for the observations. 
The typical resolution of the spectra is 2.2~{\AA} and the dispersion
2.6~{\AA} per pixel. The reduction procedure was performed with the standard routines
of the IRAF package. Examples of the reduced spectra are shown in Fig.~1.
The final results of these observations will be published elsewhere (Fox Machado et al., in preparation).
The author acknowledges the financial support from the UNAM through grant
PAPIIT IN104612 and from the IAU.


\begin{thebibliography}{}

\bibitem[Aerts et al. 2010]{aerts}
{Aerts, C., Christensen-Dalsgaard, J., \& Kurtz, D.~W.} 2010, \textit{Asteroseismology}, Springer Science+Business Media B.V.

\bibitem[Borucki et al. (2010)]{borucki}
{Borucki, W.~J., Koch, D., Basri, G., et al.} 2010,
\textit{Science}, 327, 977 

\bibitem[Fox Machado et al. (2010)]{fox}
{Fox Machado, L., Alvarez, M., Michel, R., et al.} 2010,
\textit{New Astronomy}, 15, 397 

\bibitem[Latham et al. 2005]{latham}
{Latham, D.~W., Brown, T.~M., Monet, D.~G., Everett, M., Esquerdo, G.~A., \& Hergenrother, C.~W.} 2005,
\textit{AAS}, 37, 1340

\bibitem[Molenda-\.Zakowicz et al. (2011)]{molenda}
{Molenda-\.Zakowicz, J., Latham, D.~W.,  Catanzaro, G., Frasca, A., \& Quinn, S.~N.} 2011,
\textit{MNRAS}, 412, 1210

\bibitem[Pe\~na et al. 2011]{pena}
{Pe\~na, J.~H., Fox Machado, L., Garc\'{\i}a, H., et al.} 2011,
\textit{RevMexAA}, 47, 309 

\bibitem[Uytterhoeven et al. (2011)]{uytter}
{Uytterhoeven, K., Moya, A., Grigahc\`ene, A., et al.} 2011,
\textit{A\&A}, 534, A125


\end{thebibliography}
\end{document}